# Gender Identification using MFCC for Telephone Applications – A Comparative Study

Jamil Ahmad, Mustansar Fiaz, Soon-il Kwon, Maleerat Sodanil, Bay Vo, and *Sung Wook Baik

***Abstract*—** Gender recognition is an essential component of automatic speech recognition and interactive voice response systems. Determining gender of the speaker reduces the computational burden of such systems for any further processing. Typical methods for gender recognition from speech largely depend on features extraction and classification processes. The purpose of this study is to evaluate the performance of various state-of-the-art classification methods along with tuning their parameters for helping selection of the optimal classification methods for gender recognition tasks. Five classification schemes including k-nearest neighbor, naïve Bayes, multilayer perceptron, random forest, and support vector machine are comprehensively evaluated for determination of gender from telephonic speech using the Mel-frequency cepstral coefficients. Different experiments were performed to determine the effects of training data sizes, length of the speech streams, and parameter tuning on classification performance. Results suggest that SVM is the best classifier among all the five schemes for gender recognition.

## I. Introduction

THE human voice provides semantics of the spoken words and also contains information about speaker dependent characteristics. Such speaker dependent information include speaker identity, gender, age range, and emotional state. This information can be used to build customer adaptive interactive voice response (IVR) systems. They can aid automatic human computer interaction systems to efficiently understand user needs and respond more appropriately [1]. Additionally, such information can also serve as essential analytic in decision making. Determination of human behavioral state can be useful in many applications including education, human computer interaction, and healthcare [2]. Information about the gender of a person is an important component for effective behavioral analytics. Gender recognition reduces the complexity of automatic speech recognition (ASR) and interactive voice response systems and improves their efficiency.

Identification of gender from short speech utterances has emerged as a challenging task gaining considerable attention recently. Gender recognition systems can help IVR systems in deciding appropriate responses or can aid human operators in responding in a more appropriate manner. This speaker dependent characteristic can be effectively applied to the IVR systems for Telephone calls. Some examples of the use of gender information in IVR systems include: identification of angry or unsatisfied customers [3] and adaptive waiting queue music systems [4].

Common approaches for gender recognition are based on the analysis of pitch of the speech. However, gender recognition using a single feature is not sufficiently accurate for a large variety of speakers. To capture differences in both time domain and frequency domain, a set of features known as Mel-frequency cepstrum coefficients (MFCC) are used [5]. These are widely used state-of-the-art features for automatic speech and speaker recognition. MFCC features are extracted from speech signals over a small window of 20 to 40 milliseconds. These features are also known to work efficiently in noisy environments. Due to their robust nature, they are widely used in speaker recognition tasks [6].

An essential component of gender recognition systems is the use of appropriate statistical classifier. The selection of a suitable classifier with optimized parameters is as important as the feature extraction. MFCC features from the training data are used to train the supervised classifiers. The classification models are then used to predict gender of the test subjects. Several studies have been conducted on the importance of statistical classification methods for gender recognition from speech. Hanguang Xiao [7] conducted a comparative study of three classifiers namely support vector machines (SVM), probabilistic neural network (PNN) and K-Nearest neighbor (KNN) for gender recognition using MFCC features. He reported that the accuracies of SVM were higher than PNN and KNN. He also showed that the normalization of MFCC features improved the classification performance. Metze et al. [8] performed a comparison of four different approaches towards gender recognition for telephone applications. They evaluated the performance of parallel phone recognizer, Bayesian network with prosodic features based system, linear prediction analysis system, and Gaussian mixture model (GMM) with MFCC features. The results reported that the parallel phone recognizer system was superior for long streams of speech. Bocklet et al. studied the effects of kernel types and their parameters for gender recognition using GMM super-

Jamil Ahmad, Mustansar Fiaz, and Soon-il Kwon are with the Digital Contents Research Institute, College of Electronics and Information Engineering, Sejong University, Seoul, Republic of Korea;

Maleerat Sodanil is the head of Information Technology Department, King Mongkut's University of Technology North Bangkok, Thailand.

Bay Vo is with the Faculty of Information Technology, Ho Chi Minh City University of Technology, Ho Chi Minh City, Vietnam

*Sung Wook Baik is the Director of Digital Contents Research Institute, College of Electronics and Information Engineering, Sejong University, Seoul, Republic of Korea. (Corresponding author e-mail: sbaik@sejong.ac.kr ).

vectors. They showed that the polynomial kernel with degree 1 performed the best with their dataset. The Performance of the classification model depends on several aspects ranging from data to the classification algorithm and the optimum tuning of parameters. The effect of training data sizes, data split ratios for various applications, speech stream lengths, and the effects of parameter tuning for the famous state-of-the-art classification schemes can be an interesting investigation.

In this work, we have used MFCC features from telephonic speech signals and evaluated several classification methods comprehensively to determine the best setup of classification scheme for gender recognition in telephone applications. The proposed gender recognition method will provide speaker dependent metadata for a sophisticated speaker intention and emotion analysis system for emergency calls.

## II. Materials and Methods

This section introduces the MFCC features, and the classifiers to be evaluated during this experimental study. A brief overview of the feature extraction process and the role of various classifiers is provided in the subsequent sub-sections.

### A. Speech Features Extraction

In this module, the speech signals are transformed from their natural waveform to a more compact parametric representation. The most successful way to perform this transformation is through mel-frequency cepstrum coefficients. A number of MFCC features (14-19) are extracted from telephonic speech signals which are then utilized by the classification methods. These features are extracted every 20-40 milliseconds during the telephone calls. Preprocessing of these features involve elimination of MFCC feature from the silent portion of the speech, feature normalization and mean vector computation.

MFCC features are based on the concept of frequency variation in the human ear's critical bandwidths. It uses two types of filters, linearly spaced and logarithmically spaced filters. Important speech characteristics are captured by expressing the signals in the Mel frequency scale having a linear frequency spacing below 1 kHz and a logarithmic scale above 1 kHz.

The entire audio stream is divided into equi-spaced frames. For each frame xi of the speech data, Fourier transform of xi is computed to obtain the cepstrum coefficients fi. A set of triangle filters scaled by Mel frequency is applied to obtain the MFCC features. MFCC features values are obtained as:

$$C_i = \sqrt{\frac{2}{k}} \sum_{j=1}^{k} \ln(m_j) \cos \left| \frac{\pi i (j-0.5)}{k} \right| \quad (1)$$

where k represents the number of triangle filters, $m_j$ shows the jth output of the filter, and $C_i$ is the ith MFCC component.

### B. Mean Prediction Values

MFCC features are computed every 20-40 ms which results in a very large number of feature vectors. These MFCC feature vectors are then processed by the classification schemes producing prediction probabilities for both gender classes. In order to reduce the possible effects of minor and trivial differences in these prediction values, their mean is computed over some period of time. This aggregation results in smoothing of the prediction probabilities making the results more representative and stable.

### C. Classification Schemes

In pattern recognition tasks, classification techniques tend to derive a decision boundary in the feature space based on feature similarity and number of classes. These classification algorithms construct mathematical models such that they produce a desirable set of outputs for a specified input. The model is trained using a subset of the data. This data contains valid class labels for tuning the classifier parameters to produce expected outputs for specified inputs. The performance is then validated on the test data [9]. The classifier outputs a label assigned to that class or probabilities. Some of the widely used classifiers used in this study are briefly described below.

i. Naïve Bayes

It is a widely used supervised classification scheme used in pattern recognition. It is based on Bayes 'decision theory'. Its decision principle is to select the most probable one. It is more suited for classification tasks using independent features. However, considerable performance was achieved in other cases as well [9]. A typical Naïve Bayes classifier is represented as:

$$classify(f1, f2,...fn) = \arg\max \rho(C=c) \prod_{i=1}^{n} \rho(F_i = f_i | C=c) \quad (2)$$

where n is the number of features, $f_i$ is the ith feature, $\rho$ represents conditional probability, and C represents dependent class variable having distinct outcomes c.

ii. Support Vector Machine

Support Vector Machines (SVM) were introduced by Vapnik as binary classifiers [10-12]. SVM uses a two-step classification process [13]. In the first step, a kernel function performs a low to high dimensional feature transformation. This transformation allows non-linearly separable data to be linearly separable at a higher dimension. There exist several kernels which can be used for this mapping, e.g. Polynomial kernel, radial basis function kernel, etc. are the most common ones. Secondly, it constructs a maximum margin hyperplane to draw the decision boundary between classes. The concept of maximum separation prevents misclassification of outliers making SVM a robust classification method. It is a widely used scheme in practical classification applications.

For a given set of labeled training samples $T=\{(x_i, l_i), i=1,2,...L$ where $x_i \in R^P$ and $l_i \in \{-1, 1\}$, a new test item is classified as:

$$f(x) = sign \sum_{i=1}^{L} \alpha_i . l_i . K(x_i, x) + b \quad (3)$$

Where $\alpha_i$ are the Lagrange Multipliers, b is the threshold value and K is the kernel. Support vectors are a small subset of the training samples for which $\alpha_i > 0$.

iii. Random Forest

The random forest (RF) classifier uses an ensemble approach that works in like a nearest neighbor predictor. Such schemes are based on divide-and-conquer approaches to improve classification performance. The main concept is to group several "weak learners" and form a "strong learner". The random forest utilizes a standard machine learning technique called a "decision tree", which corresponds to the weak learner in this ensemble. Random forests are usually constructed with a large number of trees ranging from tens to several hundred depending upon the classification task. The new input is run down all of the trees. The classification result is either the average or weighted average of all of the child nodes that are reached, or a majority vote for categorical variables produces the final outcome.

    iv. K-Nearest Neighbor

The K-nearest neighbor (KNN) classifier [14] is although a simple classifier yet yields good classification performance. The class label of the majority of K-nearest neighbors is assigned to a new test sample. The parameter K limits the neighborhood. Varying the number of neighbors can affect performance of the classification scheme.

    v. Multi-Layer Perceptron

Multilayer Perceptron (MLP) is a popular supervised classification scheme. An MLP can be trained using a number of different learning algorithms. In the MLP, two sigmoid functions are often used as activation functions:

$$\emptyset(y_i) = \tanh(v_i) \text{ and } \emptyset(y_i) = (1 + e^{-v_i})^{-1} \quad (4)$$

Where $y_i$ corresponds to the output of the ith neuron, and $v_i$ represents the weighted sum of input synapses.

The MLP architecture with connected synaptic weights in the hidden layers allow such networks to perform nonlinear mapping [15]. These weights and biases are computed iteratively through a general supervised algorithm known as Backpropagation.

## III. EXPERIMENTS AND RESULTS

### A. Dataset

The data was collected from different Korean male and female subjects. The dataset consist of both short speech utterances usually a few seconds in length and full telephone conversations ranging from 5 to 8 minutes in length.

### B. Experimental Setup

All experiments were performed on a standard Windows 7 desktop PC having an Intel Core i5-4670 CPU @ 3.40 GHz and 8.0 GB RAM. The feature extraction and classification processes were carried out on a Linux system in C++. Details of all the experiments along with results are discussed in the following sub-sections.

### C. Evaluation Metrics

To quantify the performance of the individual classification schemes, a standard evaluation metric known as classification accuracy is used. Let TP, TN, FP, and FN be the number of true positives, true negatives, false positives, and false negatives, respectively. The classification accuracy is computed as:

$$A_c = \frac{TP + TN}{TP + FP + FN + TN} \quad (5)$$

### D. Experiment 1: Effects of Training Data Sizes

This experiment was designed to assess the effects of the size of training data in the various classification schemes. Datasets were divided into a number of varying sized groups of 20%, 50%, and 80%. For each group, a separate classifier was trained and tested. The results are depicted in Figure 1. It can be seen that the performance improves with the increase in training data size. For a smaller train data, the classification performance was very low. This low performance of the classifiers can be attributed to the fact that the small set of training data could not provide the classifier with sufficient data to construct a stronger robust model. The accuracy of all the classifier at 20% training data in both datasets lie between 60-70% for all classifiers. When the size of training data was increased to 50%, significant improvements of 14-22% in classification performance were recorded in almost all schemes. However, a further increase in the training data size from 50% to 80% resulted in only minor improvements ranging from 3-6% only. Going further beyond this point did not yield any noticeable improvements. It is observed that once a classifier gets sufficient training data, it builds a strong robust model which can produce sufficiently good results, whereas, overloading the classifiers with data is unnecessary.

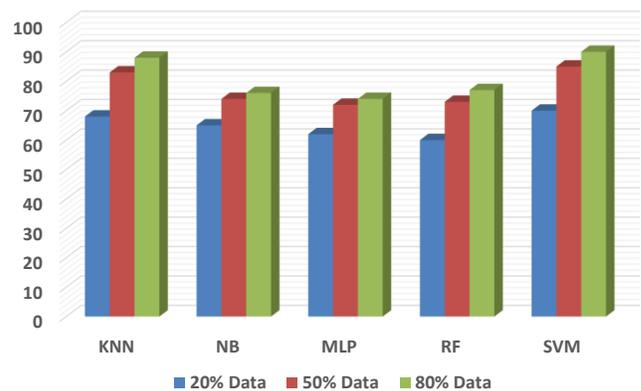

Fig. 1 Classification accuracies of various classifiers trained with varying sized training data

### E. Experiment 2: Effects of test speech stream lengths

In this experiment, it was desired to test the prediction performance of classification schemes in cases of short test streams. Classification models were trained with 80% data. Results of all the classifiers on both datasets against various test audio stream lengths are provided in Figure 2. In dataset 1, the speech streams were divided on the basis of audio lengths into 4 categories. These categories consisted to audio files of 0.5 sec, 1.0 sec, 1.5 sec, and 2.0 sec lengths. For very short audio streams, the performance of SVM, RF, and NB classifiers was better than the MLP and KNN. With the increase in the length of audio, gradual improvements were observed in the performance of all classification schemes. SVM out performed all other classifiers by a significant margin of 12-15% from its closest competitors.

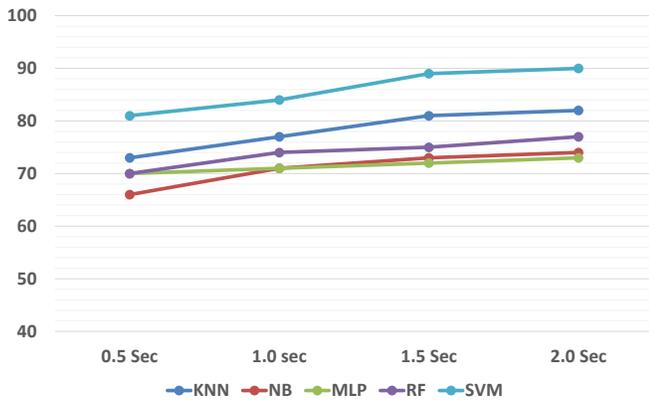

Fig. 2 Prediction performance of various classifiers with variable length test speech streams

*F. Experiment 3: Effects of Classifier Parameters*

For each classifier, there exist several parameters that can be tuned to achieve optimal performance. It is important to carefully evaluate the performance of classifiers at various configurations of these parameters. In this experiment, we evaluated the performance of all classifiers by varying the values of their respective parameters in order to determine their optimal values for maximum performance. For K-NN classifier, the number of neighbors is a parameters. In case of Naïve Bayes classifier, performance can be evaluated at normal distribution and kernel density estimation methods. In MLP, learning rate and momentum values are tested for optimality. For random forest, the objective was to determine the optimal number of decision trees. Finally, in case of SVM, various kernel configurations were evaluated.

The classification performance of all individual classifiers against various configurations are provided in tables 1 to 5. Optimal setting with maximum classification performance are highlighted in bold. In table 1, the classification accuracies against varying neighborhood sizes is given. It performs best with k=1. And its performance gradually decreases as we increase k.

TABLE I
PERFORMANCE OF KNN WITH DIFFERENT NEIGHBORHOOD SIZES

| Neighborhood size (k) | Classification Accuracy |
|---|---|
| **5** | **88** |
| 10 | 83 |
| 15 | 78 |
| 20 | 75 |

In table 2, the performance statistics of NB classifier for normal and kernel density distributions are depicted. It performs optimal with Normal distribution.

TABLE II
PERFORMANCE OF NB WITH VARYING PARAMETER VALUES

| Parameter setting | Classification Accuracy |
|---|---|
| **Normal distribution** | **76.0** |
| Kernel density estimation | 74.6 |

The classification accuracy for various learning rate and momentum values in MLP classification is provided in table 3. It can be seen that the best performance is achieved at learning rata α = 0.3 and momentum μ = 0.2. Modifying these values results in reduced classification performance.

TABLE III
PERFORMANCE OF MLP WITH VARYING PARAMETER VALUES

| Learning rate (α) | Momentum (μ) | Classification Accuracy |
|---|---|---|
| **0.3** | **0.2** | **74.2** |
| 0.3 | 0.3 | 73.4 |
| 0.4 | 0.2 | 73.1 |

The effect of increasing the number of trees in random forest is shown in table 4. It can be observed that the optimal number of trees lie in between 50 and 100 for our datasets. Performance with less than 50 trees and more than 100 trees gradually decreases.

TABLE IV
PERFORMANCE OF RANDOM FOREST WITH VARYING NUMBER OF TREES

| Number of Trees | Classification Accuracy |
|---|---|
| 10 | 75.1 |
| 20 | 76.3 |
| **50** | **77.4** |
| 100 | 76.8 |
| 200 | 76.2 |

Two commonly used kernels with varying parameter values were investigated. The parameter values are given in the brackets with each kernel type. For the polynomial kernel, the value represents degree of the polynomial. In case of RBF kernel, it corresponds to the gamma. It is shown in table 5 that the performance of both kernels is very close to each other.

TABLE V
PERFORMANCE OF SVM WITH VARIOUS KERNELS

| Kernel functions | Classification Accuracy |
|---|---|
| Polynomial kernel (1) | 87.2 |
| **Polynomial kernel (2)** | **90.1** |
| Polynomial kernel (3) | 88.3 |
| Polynomial kernel (4) | 88.0 |
| Polynomial kernel (5) | 87.2 |
| RBF kernel (0.2) | 88.2 |
| RBF kernel (0.4) | 89.4 |
| **RBF kernel (0.6)** | **90.0** |
| RBF kernel (0.8) | 89.2 |
| RBF kernel (1.0) | 89.7 |

IV. CONCLUSION

In this paper, the classification performance of five different classifiers KNN, MLP, SVM, Naïve Bayes and Random Forest were comprehensively investigated for gender recognition tasks using the widely used MFCC features. Data was collected from several Korean male and female subjects consisting of short speech utterances as well as lengthy telephone call recordings. MFCC features were extracted from the speech data at regularly spaced intervals. Silence segments were eliminated during the feature extraction phase. Three different experiments were designed to evaluate the performance of all the classifiers. In the first experiment, the objective was to investigate the effect of the size of training data. It was determined that the optimal size of training data for our telephone calls dataset lie in the range 70-85%. The performance of SVM was better than the other four classifiers.

The second experiment was an attempt to determine the performance of various classifiers in case of very short speech utterances. For very short utterance, it becomes very difficult to recognize gender because of scarce data. In this experiment, SVM performed best with very short utterances. For an increase in the length of speech, there was a gradual increase in performance of all classification schemes.

In order to determine the optimal set of parameters for the various classification schemes, several experiments were carried out. The maximum performance of each method along with the optimal configuration is provided in the results.

For all the experiments conducted, SVM with polynomial kernel of degree 2 performed better than the other four classifiers. This superior performance of SVM makes it a method of choice for gender recognition from telephonic conversation.

ACKNOWLEDGEMENT

This work was supported by the ICT R&D program of MSIP/IITP. (No. R0126-15-1119, Development of a solution for situation-awareness based on the analysis of speech and environmental sounds).